\documentclass[aps,prc,superscriptaddress,showpacs,amssymb,amsmath,amsfonts,twocolumn,floatfix]{revtex4}
\usepackage{graphicx}

\begin{document}

\include{defs}

\title{Updated SAID analysis of pion photoproduction data}

\author{Ron L.\ Workman,
	William J.\ Briscoe,
	Mark W.\ Paris,
	Igor I.\ Strakovsky}
\affiliation{Institute for Nuclear Studies, Department of Physics, 
	The George Washington University, Washington, D.C. 20052}

\date{\today}

\begin{abstract}
Energy-dependent and single-energy fits to the existing pion photoproduction
database have been updated to cover the region from threshold to 2.7~GeV in 
the laboratory photon energy. Revised resonance photo-decay couplings have 
been extracted and compared to previous determinations. The influence of 
recent measurements is displayed. Remaining problems and future approaches 
are discussed.
\end{abstract}


\maketitle

\section{Introduction}
\label{sec:intro} 

The SAID photoproduction analyses have been updated periodically since
1990~\cite{pr90}, with more frequent updates published through our 
website~\cite{said}. Our last full analysis~\cite{pr02} has been
revised twice~\cite{du1,du2} to include CLAS differential cross sections
for neutral and charged pion production off proton targets. Some recent 
neutron target data were poorly predicted by these and the MAID~\cite{maid} 
solutions, requiring changes in the neutron multipoles and resonance 
couplings. Further changes are expected with 
the incorporation of forthcoming data from JLab 
FROST~\cite{frost} and HD-ICE~\cite{hd-ice}, CB\@MAMI~\cite{mami}, 
LEPS~\cite{leps}, and CB-ELSA~\cite{cbelsa}. Here we compare with 
previous fits and consider what changes can be expected with future additions
to the database and changes in the SAID parametrization.

In Section~\ref{sec:DB} we summarize changes to the SAID database. The 
changes reflected in our multipoles are displayed in Section~\ref{sec:MA}. 
A comparison of past and recent photo-decay amplitudes, for resonances 
giving a significant contribution to pion photoproduction, is made in 
Section~\ref{sec:ResCopl}.  Finally, in Section~\ref{sec:conc}, we 
summarize our results and comment on possible changes due to further 
measurements and changes in our parametrization form.

\section{Database}
\label{sec:DB}

The most influential additions to our database have been recent measurements 
of the photon beam asymmetry $\Sigma$ for $\vec{\gamma}n\to\pi^-p$~\cite{ma10} 
and for $\vec{\gamma}n\to\pi^0n$~\cite{sa09} from GRAAL.  These include 216 
$\Sigma$ measurements of $\pi^0n$ covering E$_\gamma$=703--1475~MeV and 
$\theta$=53--164$^\circ$ plus 99 $\Sigma$ measurements of $\pi^-p$ for 
E$_\gamma$=753--1439~MeV and $\theta$=33--163$^\circ$. 

We  note that the GRAAL contribution to $\pi^0n$ has doubled the world 
database for this reaction.  Our best fit (SN11) for $\pi^0n$ and $\pi^-p$, 
reduces the initial $\chi^2$/data from 223 and 27 (for the SAID energy-dependent
solution SP09~\cite{du1}) to 
3.1 and 4.6, respectively.  In particular, this shows that the $\pi^0n$ data 
were not well predicted, based on the existing large proton-target database 
and the much smaller $\pi^-p$ dataset.

Cross section~\cite{sc10,cr11}, $\Sigma$~\cite{el09,sp10}, and double-polarized
$C_{x'}$~\cite{si11} data for $\gamma p\to\pi^0p$ have had a lesser impact. For 
this reason, in the next section we will focus mainly on the neutron target fits 
and multipoles.

\begin{figure*}[th]
\centerline{
\includegraphics[height=0.45\textwidth, angle=90]{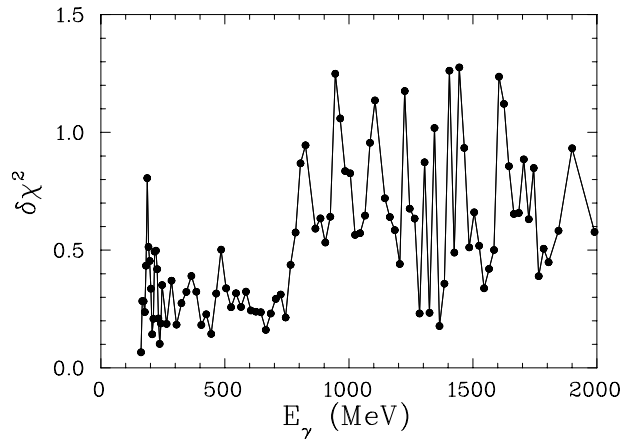}}
\vspace{3mm} \caption{Comparison of the SES and global SN11
        fits via $\delta\chi^2 =
        {[}\chi^2(SN11)-\chi^2(SES){]}$/N$_{\rm data}$ versus lab
photon energy $E_{\gamma}$. \label{fig:f1}}
\end{figure*}

\section{Multipole Amplitudes}
\label{sec:MA}

The present multipole analysis retains the phenomenological form used 
in Ref.~\cite{pr02}, 
which extended the multipole parametrization 
based on a Heitler $K$-matrix approach~\cite{pr90,km},
\begin{equation}
	M \; = \; ({\rm Born} + A)(1 + i T_{\pi N} ) + B T_{\pi N} ,
\label{eq:q1}
\end{equation}
to include a term of the form
\begin{equation}
	(C+iD)( {\rm Im} T_{\pi N} - |T_{\pi N} |^2 ), 
\label{eq:q2}
\end{equation}
where $T_{\pi N}$ is the elastic $\pi N$ scattering partial-wave amplitude 
associated with the pion-photoproduction multipole amplitude $M$.
This new piece was found necessary to fit the increasingly precise
polarization data, and has recently been used in a study of the 
model-independence of energy-dependent and single-energy 
fits~\cite{join}. The factors $A$ through $D$ were parametrized in 
terms of simple polynomials with the correct threshold behavior. 
Other forms, such as the Chew-Mandelstam (CM)
parametrization~\cite{pw},
\begin{equation}
	M \; = \; \sum_{\sigma} [1-\bar{K} C]^{-1}_{\pi\sigma} 
	\bar{K}_{\sigma\gamma} ,
\label{eq:q3}
\end{equation}
employing CM $K$-matrix elements, $\bar{K}_{\pi\sigma}$, determined in a fit 
to $\pi N$ elastic scattering data~\cite{pin}, have also been explored~\cite{pw}.

\begin{figure*}[th]
\centerline{
\includegraphics[height=0.45\textwidth, angle=90]{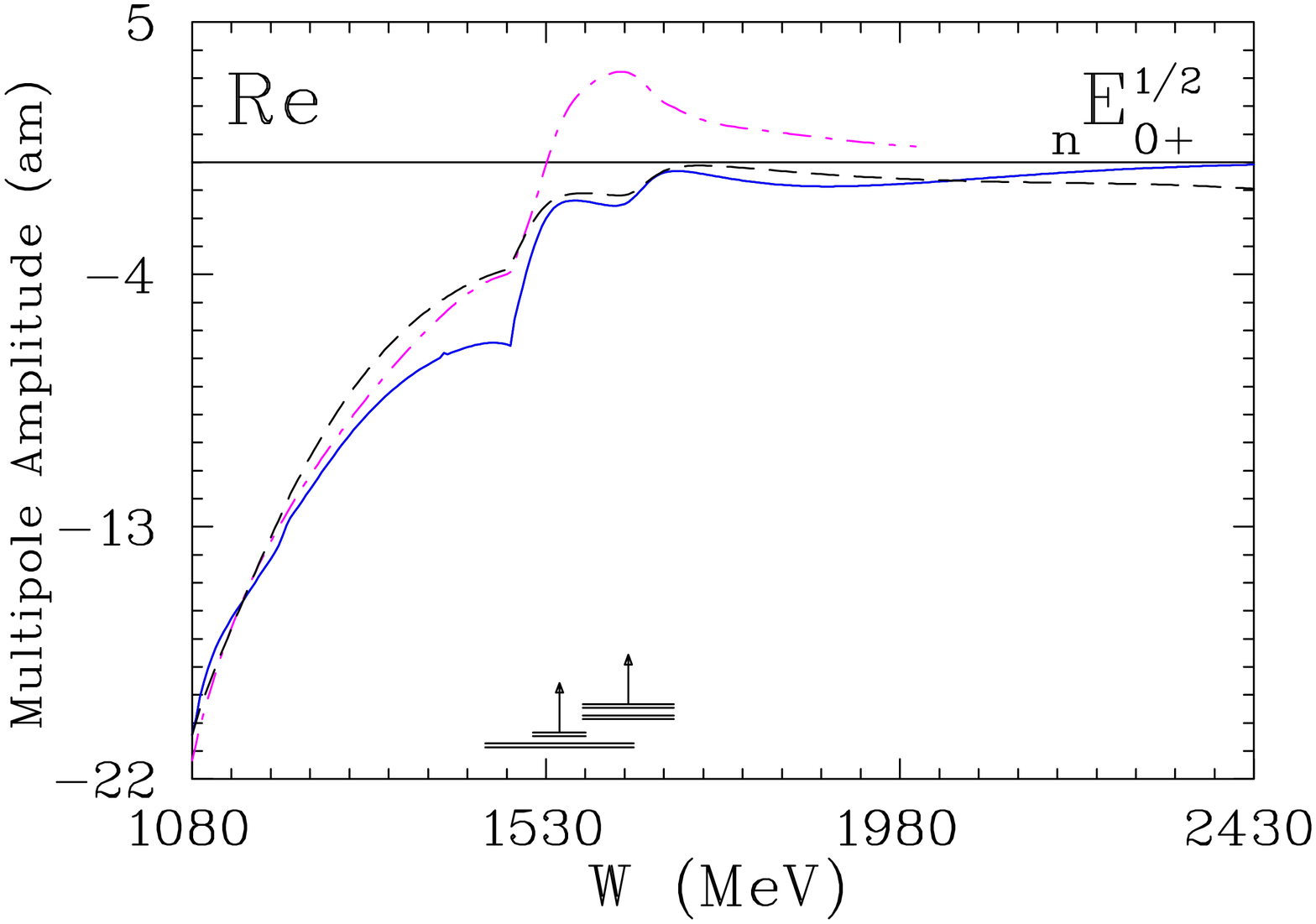}\hfill
\includegraphics[height=0.45\textwidth, angle=90]{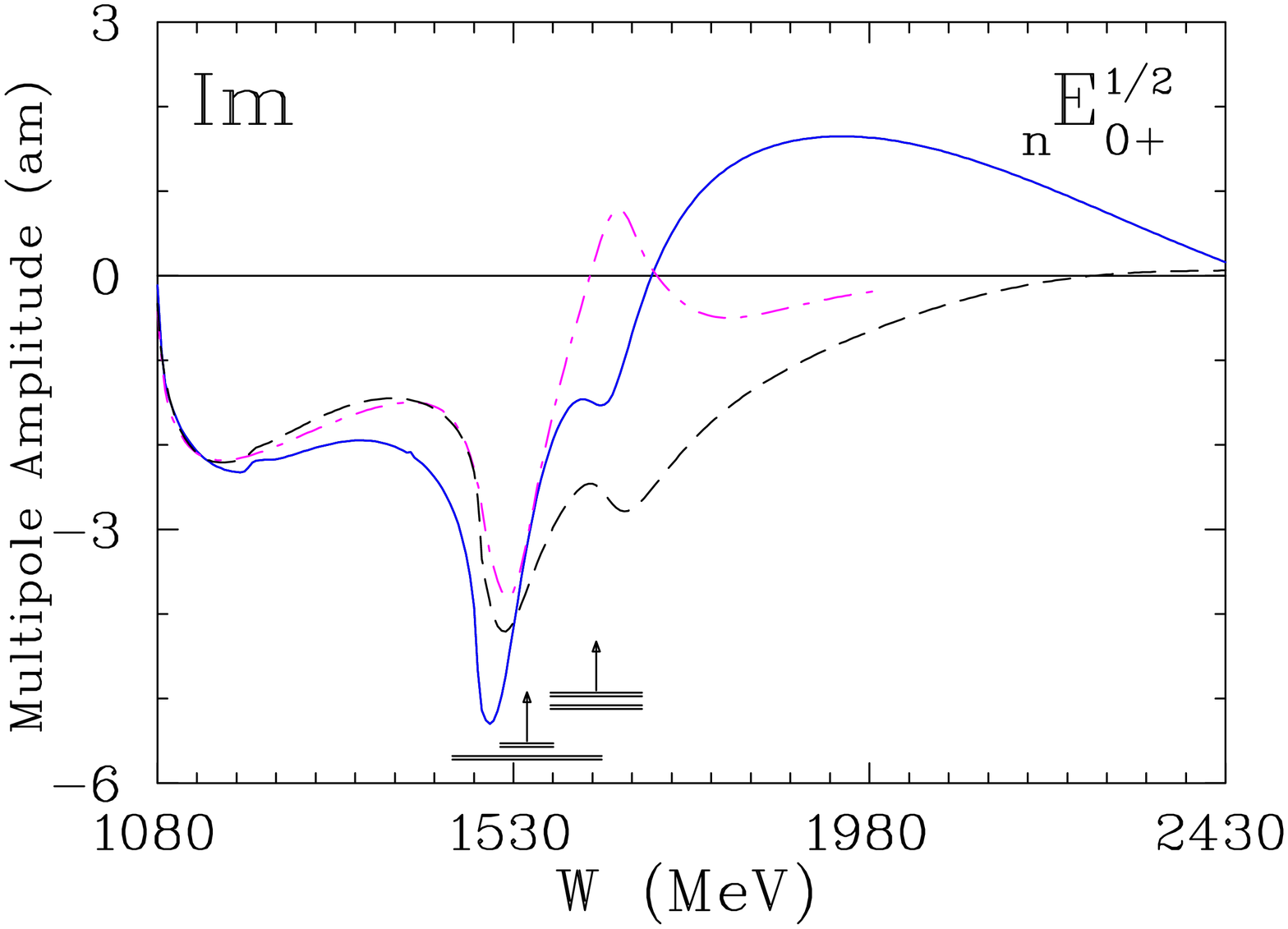}}
\centerline{
\includegraphics[height=0.45\textwidth, angle=90]{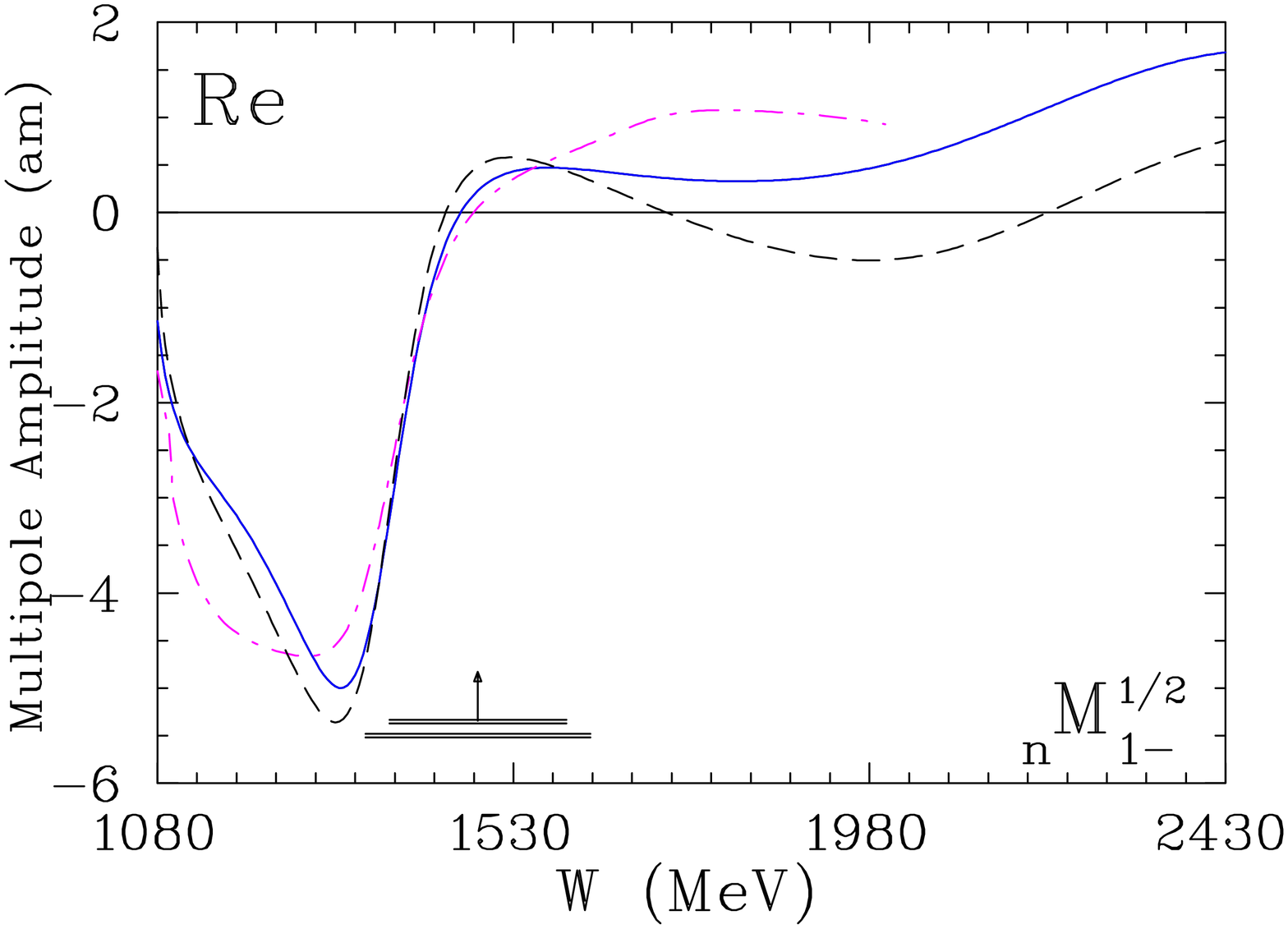}\hfill
\includegraphics[height=0.45\textwidth, angle=90]{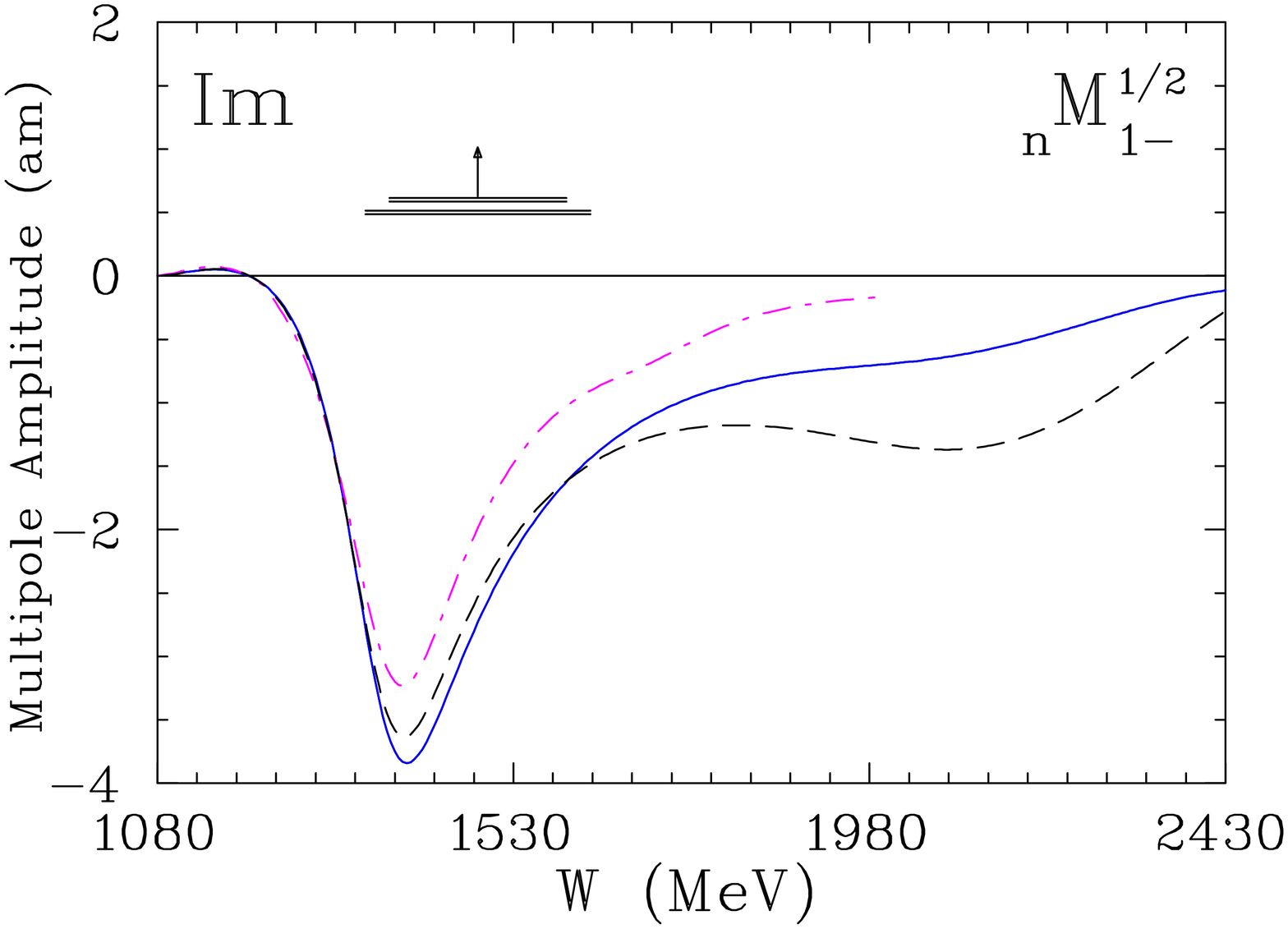}}
\centerline{
\includegraphics[height=0.45\textwidth, angle=90]{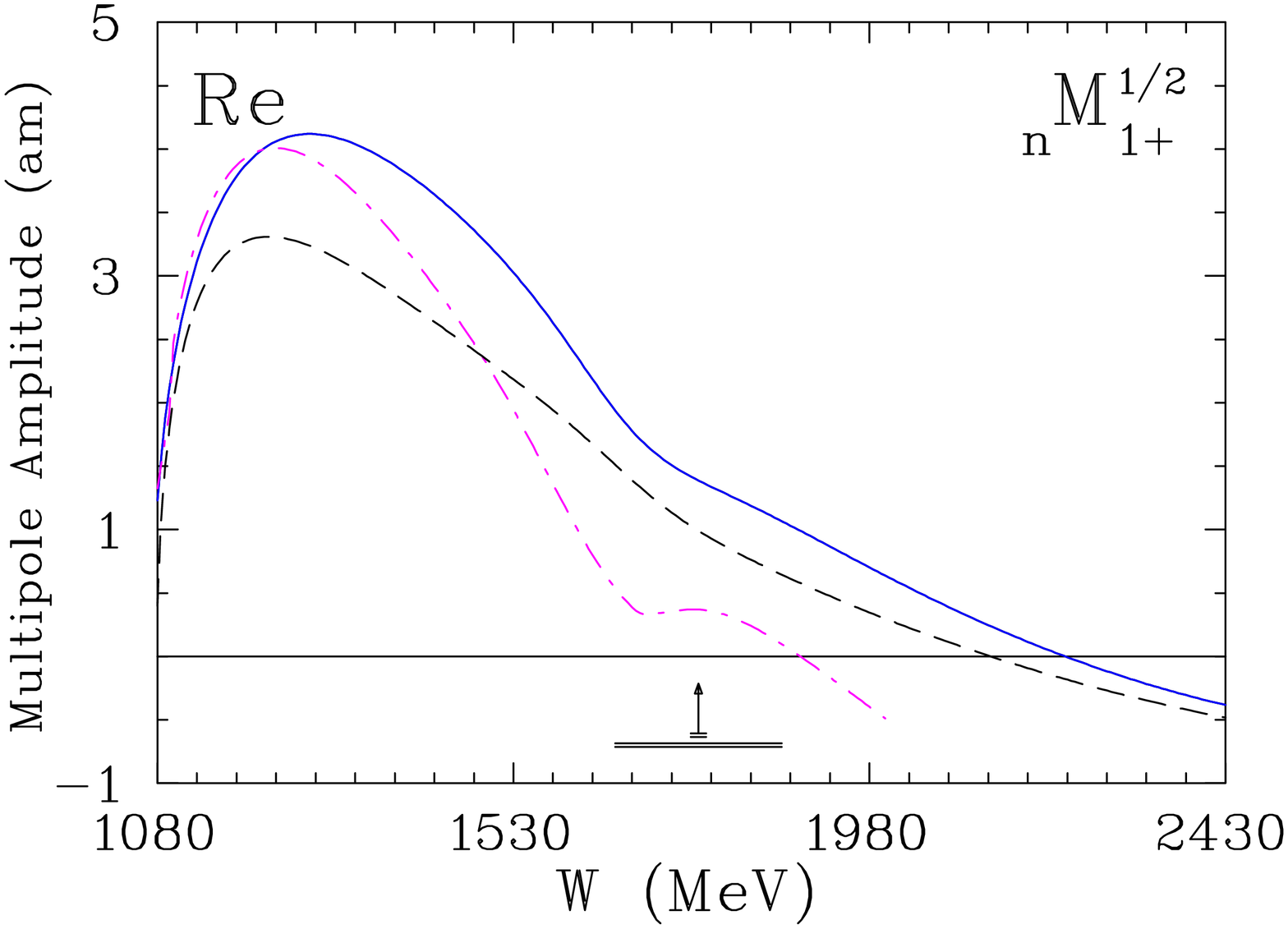}\hfill
\includegraphics[height=0.45\textwidth, angle=90]{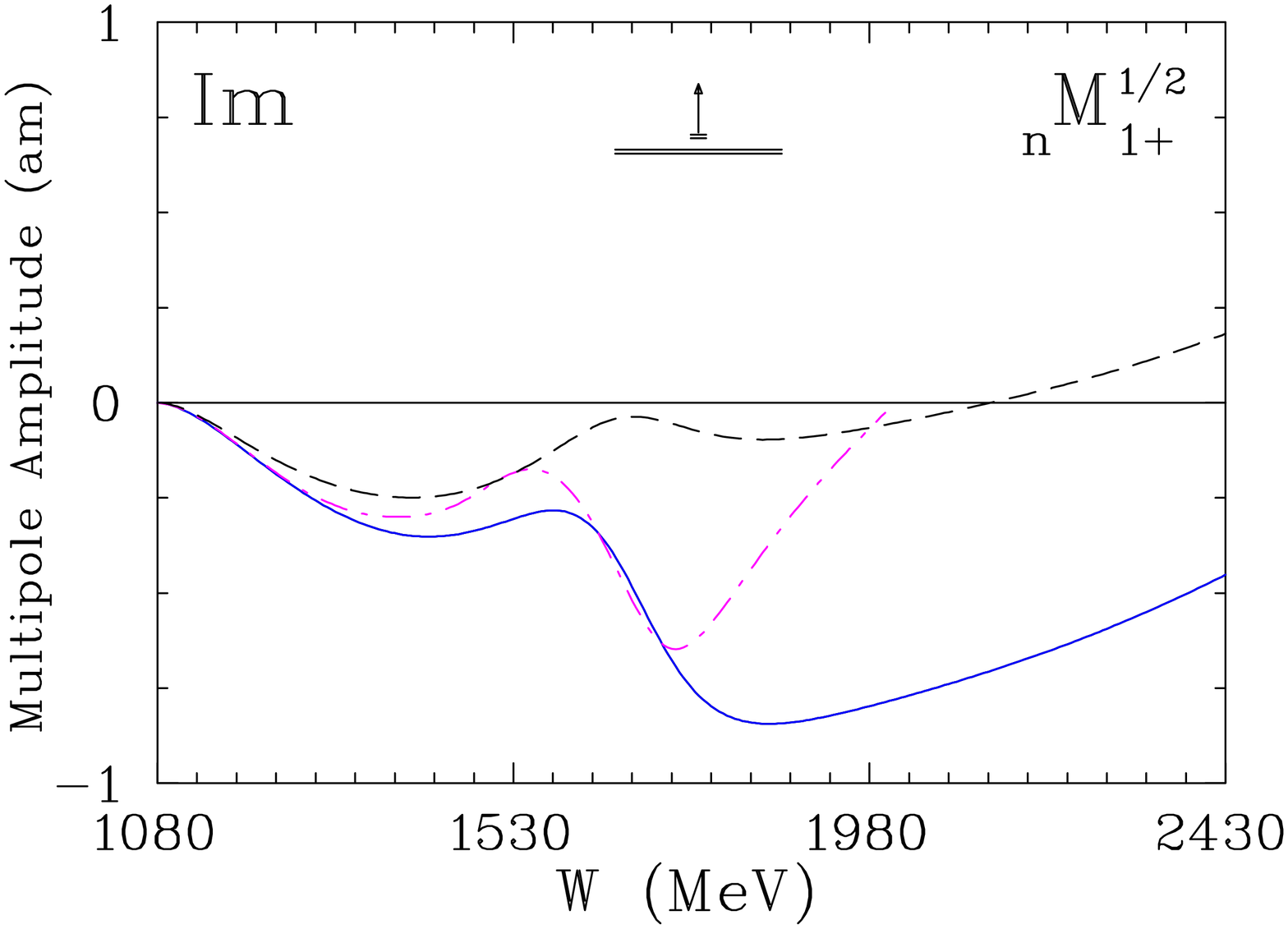}}
\centerline{
\includegraphics[height=0.45\textwidth, angle=90]{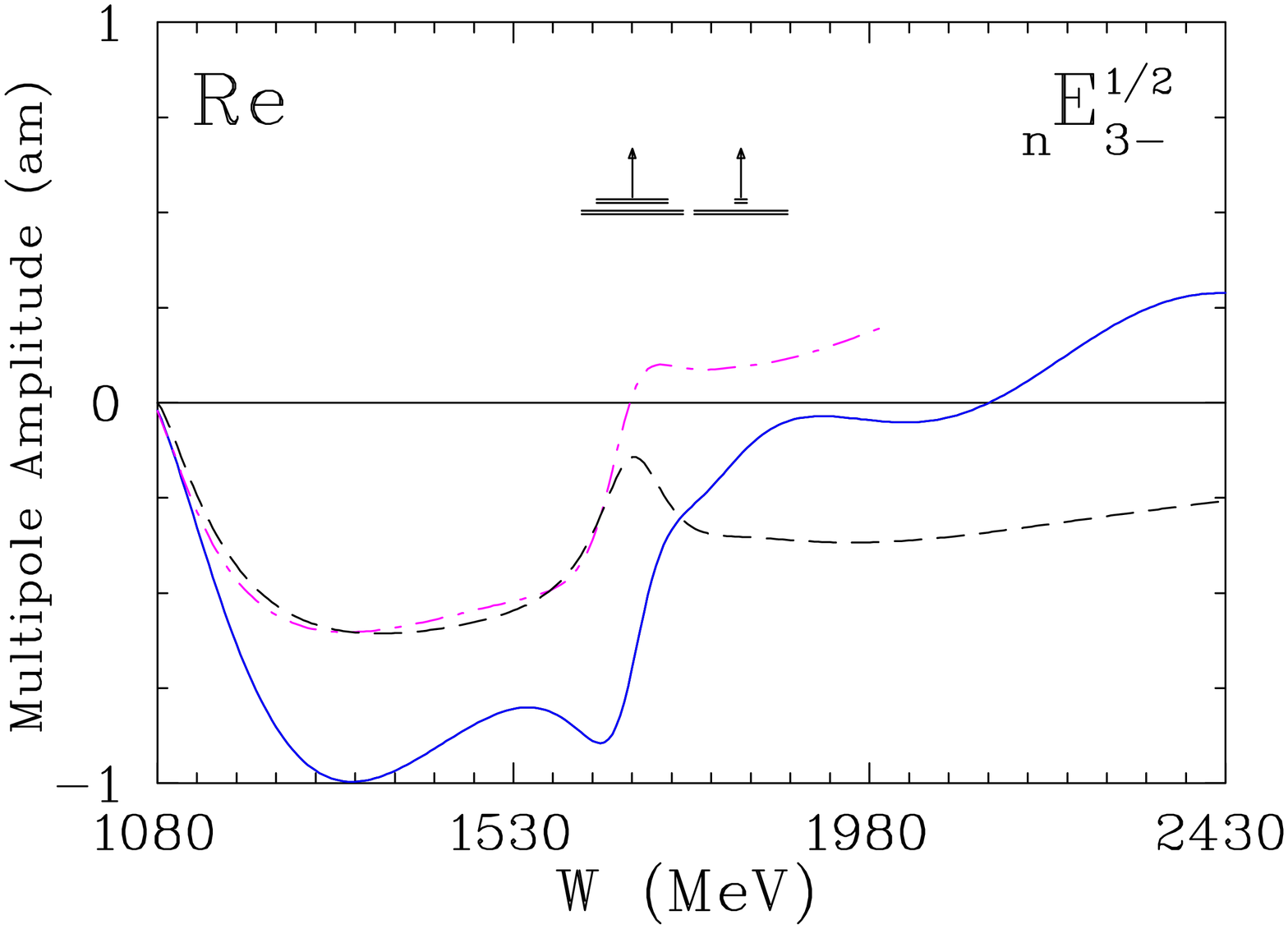}\hfill
\includegraphics[height=0.45\textwidth, angle=90]{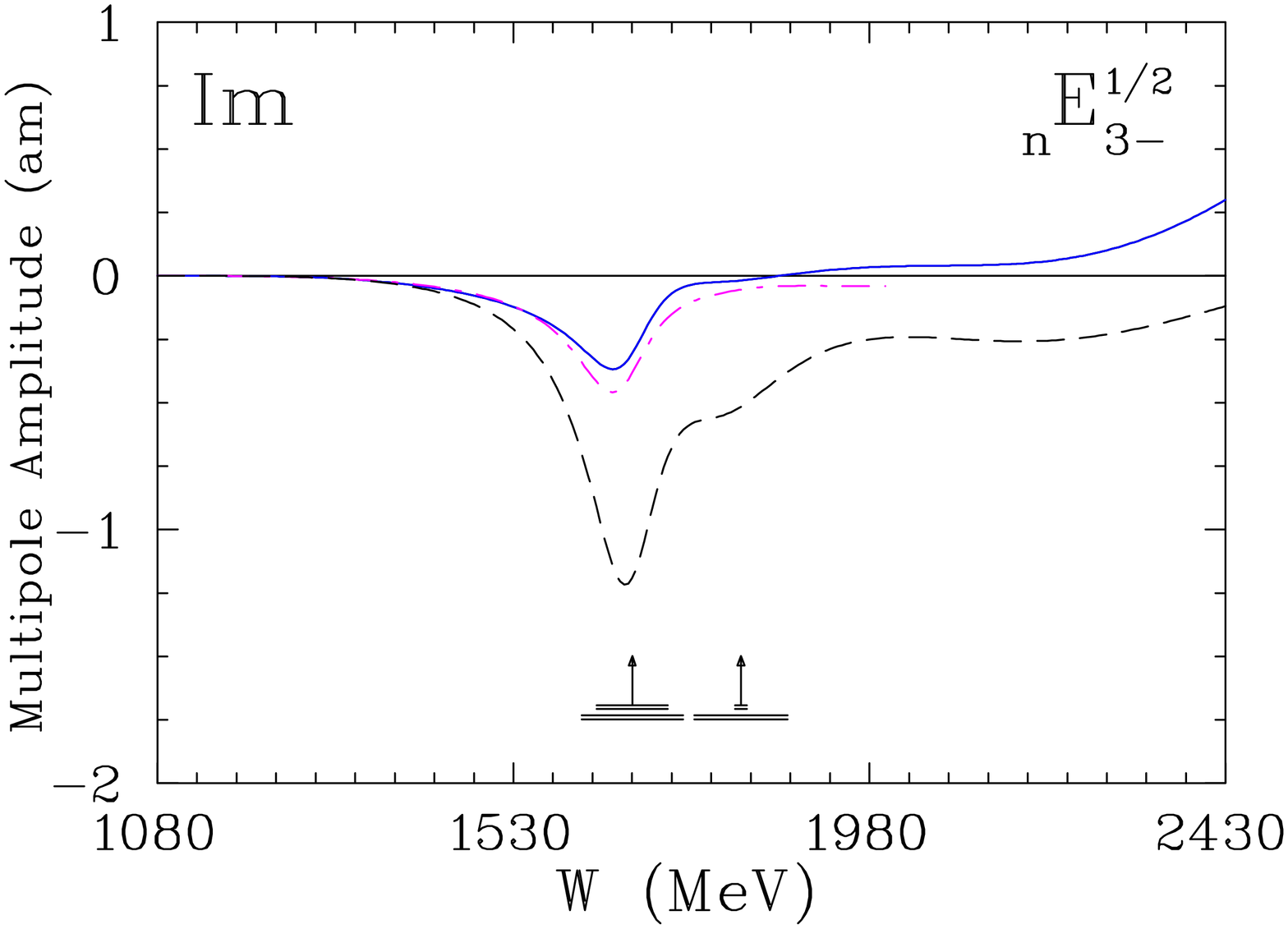}}
\caption{(Color online) Selected neutron multipole amplitudes from 
	threshold to $W$ = 2.43~GeV ($E_{\gamma}$ = 2.7~GeV). Solid lines correspond 
	to the SN11 solution.  Dashed (dash-dotted) lines give solution
	SP09~\protect\cite{du1} (MAID07~\protect\cite{maid}, which terminates at
        $W$=2~GeV).
        Vertical arrows indicate $W_R$ and horizontal bars
        show full ($\Gamma$) and partial ($\Gamma_{\pi N}$) widths
        associated with the SAID $\pi N$ solution SP06~\protect\cite{pin}.
        \label{fig:f2}}
\end{figure*}
\begin{figure*}[th]
\centerline{
\includegraphics[height=0.95\textwidth, angle=90]{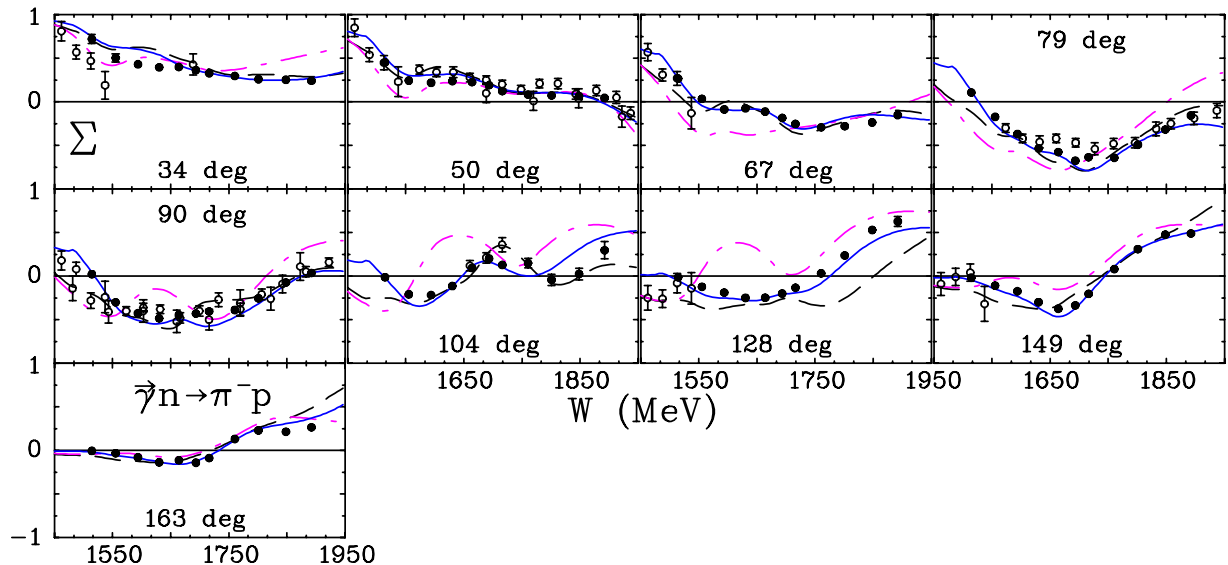}}
\vspace{3mm} \caption{(Color online) $\Sigma$ asymmetry for
        $\vec{\gamma}n\to\pi^-p$. Data (filled circles) are
        from GRAAL~\protect\cite{ma10}. The previous measurements
        (open circles) are available in the SAID
        database~\protect\cite{said}. Notation of the solutions
        is the same as in Fig.~\protect\ref{fig:f2}. GRAAL
        measurements are not included in SP09 and MAID07.
        \label{fig:f3}}
\end{figure*}
\begin{figure*}[th]
\centerline{
\includegraphics[height=0.95\textwidth, angle=90]{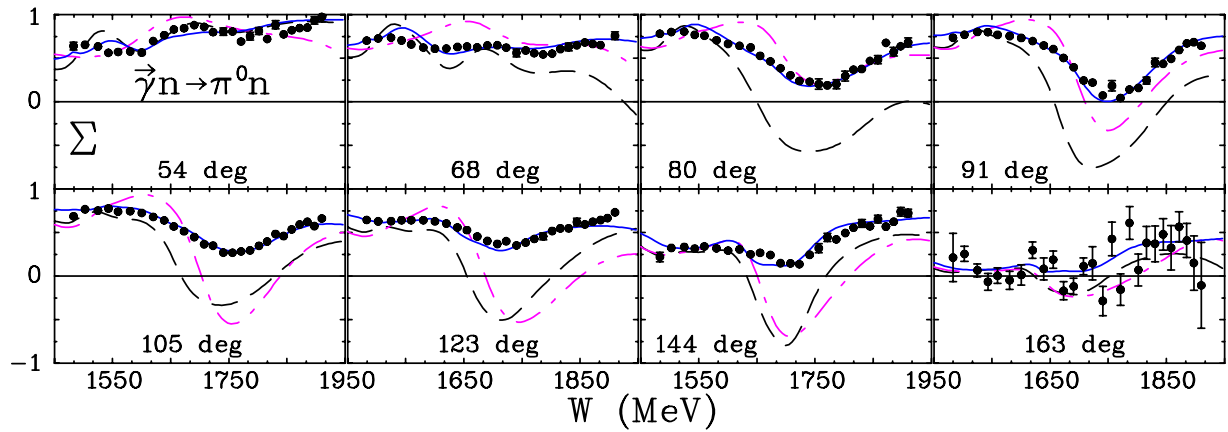}}
\vspace{3mm} \caption{(Color online) $\Sigma$ asymmetry for
        $\vec{\gamma}n\to\pi^0n$. Data (filled circles) are
        from GRAAL~\protect\cite{sa09}. Notation of the
        solutions is the same as in Fig.~\protect\ref{fig:f2}.
        GRAAL measurements not included in SP09 and MAID07.
        \label{fig:f4}}
\end{figure*}
\begin{figure*}[th]
\centerline{
\includegraphics[height=0.4\textwidth, angle=90]{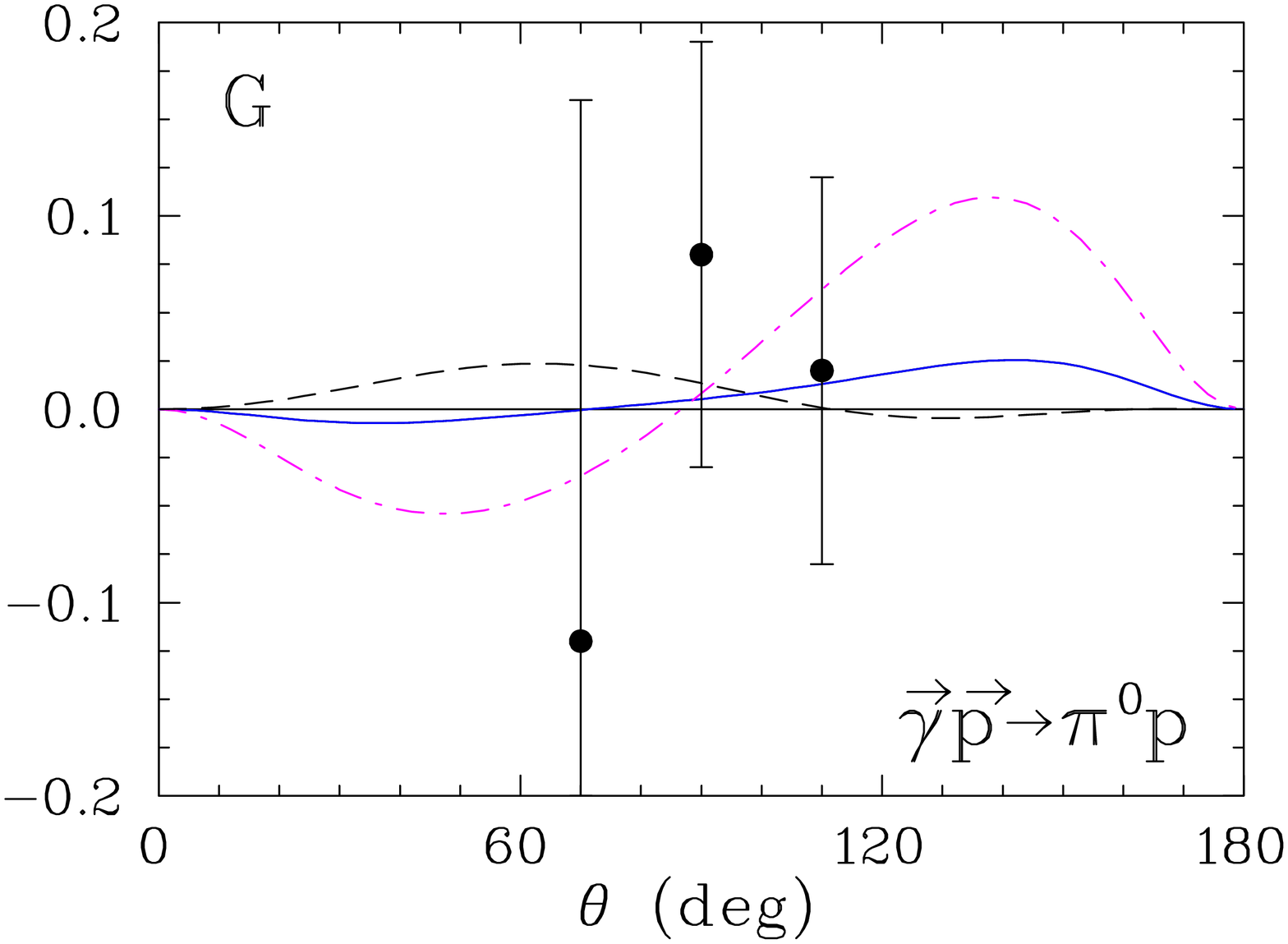}\hfill
\includegraphics[height=0.4\textwidth, angle=90]{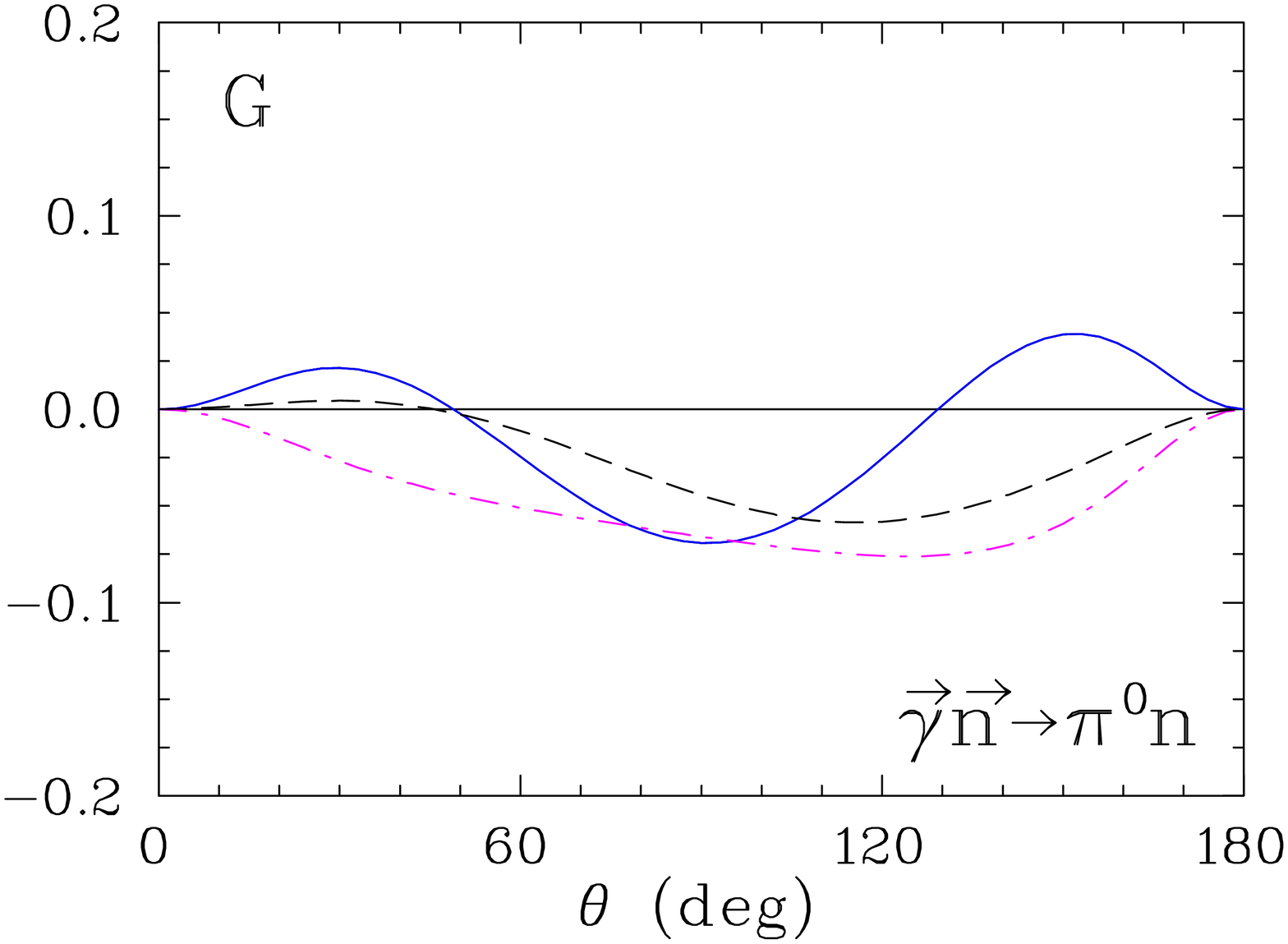}}
\vspace{3mm} \caption{(Color online) G asymmetry for neutral-pion
        photoproduction in the $\Delta$ resonance region (E$_\gamma$=340~MeV).
        Data (filled circles) are from MAMI~\protect\cite{ah05}
        Notation of the solutions is the same as in
        Fig.~\protect\ref{fig:f2}. \label{fig:f5}}
\end{figure*}
\begin{figure*}[th]
\centerline{
\includegraphics[height=0.6\textwidth, angle=90]{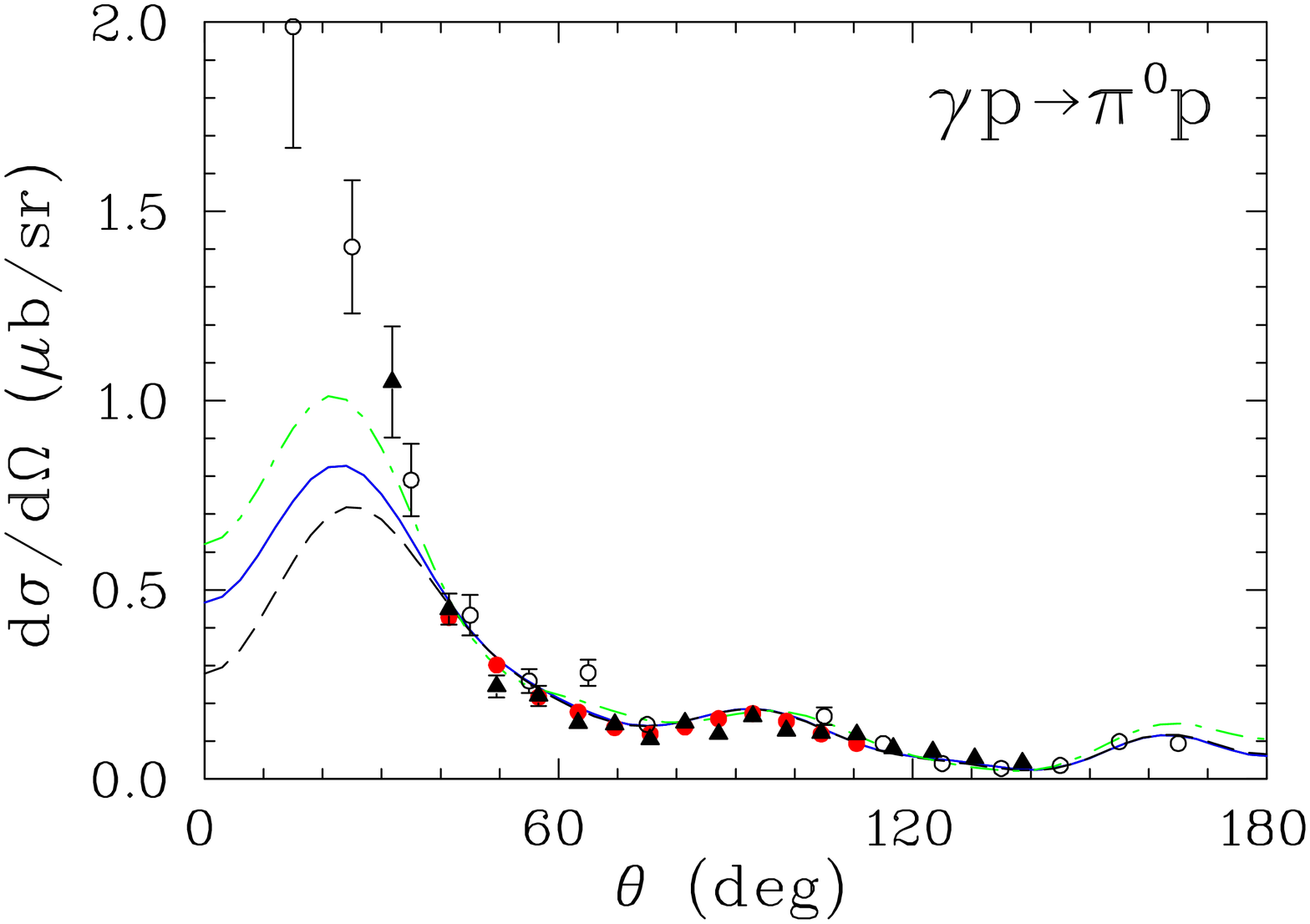}}
\vspace{3mm} \caption{(Color online) Differential cross section
        for $\gamma p\to\pi^0p$ at E$_\gamma$=2225~MeV.  CLAS data (filled
        circles) are from ~\protect\cite{du2}.  CB-ELSA data
        (open circles~\protect\cite{cr11} and filled triangles
        \protect\cite{ba05}). Notation of the solutions
        is the same as in Fig.~\protect\ref{fig:f2}. SZ11
        solution (CLAS $\pi^0p$ cross sections~\protect\cite{du1}
        excluded) is shown by a dash-dotted line. \label{fig:f6}}
\end{figure*}

The multipole amplitudes are presented in terms of isospin states, as 
is the convention. Extending such an analysis below the $\pi^+n$ 
threshold is clearly problematic. This region is not the 
focus of the present study, and requires a separate analysis. In fits 
after SM95~\cite{sm95}, Arndt proposed a recipe whereby the above 
$\pi N$ partial-wave $T$-matrices were evaluated in terms of the outgoing 
pion energy of a corresponding photoproduction reaction rather than the 
center-of-mass energy. This method allowed rather good fits to the 
threshold data, but resulted in a charge-state-dependent shift of 
the $\pi N$ $T$-matrix pole 
positions, by a few MeV, depending on whether the $\pi^0p$ or $\pi^+n$
final states were being analyzed. Here we have made fits, with (SK11) 
and without (SN11) this kinematic shift, to gauge its influence on the 
photo-decay amplitudes. The fit quality, in terms of chi-squared, for 
these and previous SAID solutions, is compared in Table~\ref{tab:tbl1}.

\begin{table}[th]
\caption{$\chi^2$ comparison of fits to pion photoproduction
        data.  Results are shown for eight different SAID solutions 
	(current SN11 and SK11 with previous SP09~\protect\cite{du1},
	FA06~\protect\cite{du2}, SM02~\protect\cite{pr02}, and
	SM95~\protect\cite{sm95}).
	See text for details. \label{tab:tbl1}}
\vspace{2mm}
\begin{tabular}{|c|c|c|c|}
\colrule
Solution & Energy limit& $\chi^2$/N$_{\rm Data}$ & N$_{\rm Data}$ \\
         & (MeV)       &               &      \\
\colrule
SN11     & 2700        &  2.08         & 25553 \\
SK11     & 2700        &  2.09         & 25961 \\
SP09     & 2700        &  2.11         & 25639 \\
SM02     & 2000        &  2.01         & 17571 \\
SM95     & 2000        &  2.37         & 13415 \\
\colrule
\end{tabular}
\end{table}

The SK11 fit extends down to the $\pi^0p$ threshold, at a photon energy
of E$_\gamma$=144.7~MeV, while SN11 is
limited to 155~MeV, just above the $\pi^+n$ threshold (151.4~MeV), thus avoiding
complications from the region between the $\pi^0p$ and $\pi^+n$ thresholds. 
The quality of the overall data fit, 
as shown in Table~\ref{tab:tbl1}, is nearly identical. 

We have generated the single-energy solutions (SES), described extensively 
in Ref.~\cite{join}, based on the global fit SN11. The quantity $\delta\chi^2$ 
is the difference, ${[}\chi^2(SN11)-\chi^2(SES){]}$, divided by the number of 
data in each single-energy bin, providing a measure of the agreement between an 
individual SES and the global SN11 results (see Fig.~\ref{fig:f1}). We 
emphasize that the SES are generated mainly to search for missing structures 
in the global fit. Detailed comparisons of the global and SES fits can be 
made on the SAID website~\cite{said}.

In Fig.~\ref{fig:f2}, we display the most significant deviations of the
SN11 solution from the fit SP09, published in Ref.~\cite{du1}, and the 
Mainz MAID07~\cite{maid} result for selected neutron multipoles.

The differences between our SN11 and SP09 results for neutron targets 
are visible particularly for the $E^{1/2}_{0+}$ (Fig.~\ref{fig:f2})
above $W \approx 1280$~MeV (E$_\gamma$=400~MeV). 
The $E_{3-}^{1/2}$ multipole, connected to the
$N(1680)F_{15}$ resonance, is also quite different. ($N(1680)$ being the PDG
notation~\cite{PDG} and $L_{2I,2J}$ being the associated notation for a state
in $\pi N$ elastic scattering~\cite{pin}.) The MAID07 fit was also 
modified~\cite{sa09} to accommodate the new $\pi^0n$ $\Sigma$ data, 
resulting in changes beginning at a higher energy, mainly altering the 
$N(1720)P_{13}$ resonance parameters. Fits to these neutron target data
are displayed in Figs.~\ref{fig:f3} and \ref{fig:f4}.

In Figs.~\ref{fig:f5} and \ref{fig:f6}, we show discrepancies corresponding to 
the double-polarization quantity ($G$), at low energy, and others for the 
unpolarized cross section, at higher energies. In Fig.~\ref{fig:f6}, we compare a 
prediction from SP09 to a fit including the data of Ref.~\cite{cr11}. A large 
discrepancy exists in the forward direction. Eliminating the existing CLAS cross 
sections~\cite{du2}, which do not extend to very forward angles, allows a slightly 
improved fit, but clearly does not resolve this problem, which exists also for the 
Bonn-Gatchina fits~\cite{boga}.  

\section{Resonance Couplings}
\label{sec:ResCopl}

In order to make meaningful comparisons~\cite{note} with previous resonance 
determinations, we have retained the method used in Refs.~\cite{pr02,du1,du2}, 
fitting the resultant multipoles with a background plus resonance 
assumption, similar to that used in the MAID analysis,
\begin{equation}
	A(W)(1 + i T_{\pi N} ) + T_{\rm BW}e^{i\phi},
\label{eq:q4}
\end{equation}
wherein $T_{\pi N}$ is again the corresponding $\pi N$ $T$-matrix and
$T_{\rm BW}$ is a Breit-Wigner (BW) parametrization of the resonance 
contribution. $A(W)$ is a linear approximation to the energy dependence of
the Born plus phenomenological term multiplying $(1+iT_{\pi N} )$ in Eq.~(\ref{eq:q1}).
Here the (theoretical) systematic error in our determination is generally
much larger than the statistical errors found in fitting the data over an
energy bin, around the BW resonance energy, or fitting (with subjective
errors) the energy-dependent or single-energy multipoles covering the same 
energy range. The errors quoted in Tables~\ref{tab:tbl2} and \ref{tab:tbl3} 
were found by varying the energy range of the fit between the estimated 
resonance full and half-widths previously determined from our $\pi N$ 
elastic scattering analysis~\cite{pin}. 

The amplitudes $A_{1/2}$ and $A_{3/2}$ were determined assuming the 
masses, widths, and $\pi N$ branching fractions determined in an 
earlier BW analysis of $\pi N$ elastic scattering data~\cite{pin}. 
In a few cases, these input parameters from $\pi N$ scattering did 
not produce good fits to the photoproduction multipoles.

For the $N(1650)S_{11}$, increasing the mass to the nominal value of 
1650~MeV produced a much better fit, as did increasing the width. 
The photo-decay couplings (proton and neutron) changed substantially 
and, as a result, this variability was taken to determine the quoted 
errors. The $N(1535)S_{11}$ decay to $p\gamma$ has remained quite stable 
while the decay to $n\gamma$ has changed significantly, due mainly 
to the new $\Sigma$ data for both the $\pi^-p$ and $\pi^0n$ channels, 
as shown in terms of the multipoles and data fits in 
Figs.~\ref{fig:f2}--\ref{fig:f4}. 

The $N(1720)P_{13}$ neutron couplings are poorly determined, but the SN11 
solution has, 
for the neutron $M^{1/2}_{1+}$ multipole, an imaginary part now more closely 
resembling the MAID07 value at the BW resonance energy. The MAID07 values 
for the 
neutron $A_{1/2}$ and $A_{3/2}$ amplitudes are $-3$ and $-31$~GeV$^{-1/2} 
\times 10^{-3}$, in better agreement with SAID, compared to the last 
published values from the SM95 fit. The $\Delta (1620)S_{31}$ 
amplitude is now significantly larger, and outside 
of the PDG estimate, but is consistent with the MAID07~\cite{maid} and 
Bonn-Gatchina~\cite{boga} results (66 and 63$\pm$12~GeV$^{-1/2} \times 
10^{-3}$, respectively).

\section{Summary and Conclusion}
\label{sec:conc}

This updated analysis examined mainly the effect of new
neutron-target data on the SAID multipoles and resonance
parameters. In some cases, the changes have been significant.
The neutron multipoles generally show much larger variations
than the proton multipoles, when the fits of different 
groups are compared. Given the inability of fits, based 
mainly on proton target data, to predict the $\pi^0n$
multipoles, further changes can be expected as more 
neutron-target data become available. 

Apart from a few special cases,
the photo-decay amplitudes, $A_{1/2}$ and $A_{3/2}$, found
in this study are reasonably consistent with the PDG averages.
The $N(1535)S_{11}$ and $N(1650)S_{11}$ amplitudes deserve further discussion, though
they are now also consistent with the PDG estimates. The large PDG uncertainty,
assigned to the $N(1535)S_{11}$, was due mainly to a disagreement
which existed between values determined from eta photoproduction fits,
and existing values from pion photoproduction. Roughly, in 1995, the eta 
photoproduction value for $A_{1/2}$~\cite{krusche95} was twice the SAID SM95 
value from pion photoproduction. While the SAID value has migrated up to a 
value consistent with the early eta photoproduction estimates, the MAID 
determination has decreased, once again leaving a wide discrepancy. The 
increased SAID value for the $N(1650)S_{11}$ amplitude appears to be due more to 
the extraction technique than any change in the multipole. As we mentioned 
above, this extraction was very sensitive to assumed values for the mass 
and width, which may have produced the low value in Ref.~\cite{du2}.

The use of Eq.~(\ref{eq:q4}) in extracting the above values of $A_{1/2}$ 
and $A_{3/2}$ is reasonably consistent with the MAID approach and earlier 
extractions by Berends and Donnachie~\cite{bd} and by Crawford and 
Morton~\cite{crawford}. It may not, however, be consistent with 
determinations~\cite{boga} based on the $K$- or $T$-matrix pole. This complicates 
comparisons, particularly for multipoles without clear resonance signatures. 
We hope to address this point in future studies. 

Finally, we mention that preliminary fits to photoproduction data, using the
CM formalism of Eq.~(\ref{eq:q3}), discussed in Ref.~\cite{pw}, are qualitatively 
similar to, but quantitatively different from, the results presented here. This form, 
which uses a more constrained approach to the incorporation of higher opening 
channels ($\eta N$, $\pi \Delta$, $\rho N$), essentially replaces the behavior 
of the term given in Eq.~(\ref{eq:q2}) (proportional to the reaction cross 
section) by terms contributing to each channel separately. A more detailed 
comparison is in progress.

\begin{acknowledgments}
This work was supported in part by the U.S.\ Department of Energy
Grant DE--FG02--99ER41110. 
\end{acknowledgments}


\begin{table*}[th]
\caption{Resonance parameters for N$^\ast$ and $\Delta^\ast$ states
         from the SAID fit to the $\pi N$ data~\protect\cite{pin}
         (second column) and proton helicity amplitudes $A_{1/2}$ and
         $A_{3/2}$ (in [(GeV)$^{-1/2}\times 10^{-3}$] units) from the SN11
         solution (first row), previous SP09~\protect\cite{du1}
         solution (second row), and average values from the
         PDG10~\protect\cite{PDG} (third row). \label{tab:tbl2}}
\vspace{2mm}
\begin{tabular}{|c|c|c|c|}
\colrule
Resonance        & $\pi N$ SAID               &   $A_{1/2}$    & $A_{3/2}$ \\
\colrule
$N(1535)S_{11}$  & $W_{R}$=1547~MeV           &   99$\pm$2     & \\
                 & $\Gamma$=188~MeV           &100.9$\pm$3.0   & \\
                 & $\Gamma _{\pi}/\Gamma$=0.36&   90$\pm$30    & \\
\colrule
$N(1650)S_{11}$  & $W_{R}$=1635~MeV           &   65$\pm$25    & \\
                 & $\Gamma$=115~MeV           &  9.0$\pm$9.1   & \\
                 & $\Gamma _{\pi}/\Gamma$=1.00&   53$\pm$16    & \\
\colrule
$N(1440)P_{11}$  & $W_{R}$=1485~MeV           &$-$58$\pm$1     & \\
                 & $\Gamma$=284~MeV           &$-$56.4$\pm$1.7 & \\
                 & $\Gamma _{\pi}/\Gamma$=0.79&$-$65$\pm$4     & \\
\colrule
$N(1720)P_{13}$  & $W_{R}$=1764~MeV          &   99$\pm$3     & $-$43$\pm$2      \\
                 & $\Gamma$=210~MeV          &   90.5$\pm$3.3 & $-$36.0$\pm$3.9  \\
                 & $\Gamma _{\pi}/\Gamma$=0.09&  18$\pm$30    & $-$19$\pm$20     \\
\colrule
$N(1520)D_{13}$  & $W_{R}$=1515~MeV          &$-$16$\pm$2     & 156$\pm$2        \\
                 & $\Gamma$=104~MeV          &$-$26.0$\pm$1.5 & 141.2$\pm$1.7    \\
                 & $\Gamma _{\pi}/\Gamma$=0.63&$-$24$\pm$9    & 166$\pm$5        \\
\colrule
$N(1675)D_{15}$  & $W_{R}$=1674~MeV          &   13$\pm$2     &  19$\pm$2        \\
                 & $\Gamma$=147~MeV          &   14.9$\pm$2.1 &  18.4$\pm$2.1    \\
                 & $\Gamma _{\pi}/\Gamma$=0.39&  19$\pm$8     &  15$\pm$9        \\
\colrule
$N(1680)F_{15}$  & $W_{R}$=1680~MeV          &$-$13$\pm$3     & 141$\pm$3        \\
                 & $\Gamma$=128~MeV          &$-$17.6$\pm$1.5 & 134.2$\pm$1.6    \\
                 & $\Gamma _{\pi}/\Gamma$=0.70&$-$15$\pm$6    & 133$\pm$12       \\
\colrule
\colrule
$\Delta(1620)S_{31}$& $W_{R}$=1615~MeV       &   64$\pm$2     & \\
                 & $\Gamma$=147~MeV          &   47.2$\pm$2.3 & \\
                 & $\Gamma _{\pi}/\Gamma$=0.32&  27$\pm$11    & \\
\colrule
$\Delta(1232)P_{33}$& $W_{R}$=1233~MeV       &$-$138$\pm$3    & $-$259$\pm$5    \\
                 & $\Gamma$=119~MeV          &$-$139.6$\pm$1.8& $-$258.9$\pm$2.3\\
                 & $\Gamma _{\pi}/\Gamma$=1.00&$-$135$\pm$6   & $-$250$\pm$8    \\
\colrule
$\Delta(1700)D_{33}$& $W_{R}$=1695~MeV       &  109$\pm$4     & 84$\pm$2        \\
                 & $\Gamma$=376~MeV          &  118.3$\pm$3.3 & 110.0$\pm$3.5   \\
                 & $\Gamma _{\pi}/\Gamma$=0.16& 104$\pm$15    &  85$\pm$22      \\
\colrule
$\Delta(1905)F_{35}$& $W_{R}$=1858~MeV       &    9$\pm$3   & $-$46$\pm$3       \\
                 & $\Gamma$=321~MeV          &   11.4$\pm$8.0 & $-$51.0$\pm$8.0 \\
                 & $\Gamma _{\pi}/\Gamma$=0.12&   26$\pm$11   & $-$45$\pm$20    \\
\colrule
$\Delta(1950)F_{37}$& $W_{R}$=1921~MeV       & $-$71$\pm$2    & $-$92$\pm$2     \\
                 & $\Gamma$=271~MeV          & $-$71.5$\pm$1.8& $-$94.7$\pm$1.8 \\
                 & $\Gamma _{\pi}/\Gamma$=0.47& $-$76$\pm$12  & $-$97$\pm$10    \\
\colrule
\end{tabular}
\end{table*}

\begin{table*}[th]
\caption{Resonance parameters for N$^\ast$ states
         from the SAID fit to the $\pi N$ data ~\protect\cite{pin}
         (second column) and neutron helicity amplitudes $A_{1/2}$ and
         $A_{3/2}$ (in [(GeV)$^{-1/2} \times 10^{-3}$] units) from the SN11
         solution (first row), previous SM02~\protect\cite{pr02}
         solution (second row), and average values from the
         PDG10~\protect\cite{PDG} (third row). $^{\dagger}$SM95
         value~\protect\cite{sm95} \label{tab:tbl3}}
\vspace{2mm}
\begin{tabular}{|c|c|c|c|}
\colrule
Resonance        & $\pi N$ SAID              &   $A_{1/2}$   & $A_{3/2}$   \\
\colrule
$N(1535)S_{11}$  & $W_{R}$=1547~MeV          &  $-$60$\pm$3  & \\
                 & $\Gamma$=188~MeV          &  $-$16$\pm$5  & \\
                 & $\Gamma _{\pi}/\Gamma$=0.36&  $-$46$\pm$27& \\
\colrule
$N(1650)S_{11}$  & $W_{R}$=1635~MeV          &  $-$26$\pm$8  &  \\
                 & $\Gamma$=115~MeV          &  $-$28$\pm$4  & \\
                 & $\Gamma _{\pi}/\Gamma$=1.00& $-$15$\pm$21 & \\
\colrule
$N(1440)P_{11}$  & $W_{R}$=1485~MeV          &   48$\pm$4    & \\
                 & $\Gamma$=284~MeV          &   45$\pm$15   & \\
                 & $\Gamma _{\pi}/\Gamma$=0.79&  40$\pm$10   & \\
\colrule
$N(1720)P_{13}$  & $W_{R}$=1764~MeV          & $-$21$\pm$4   & $-$38$\pm$7  \\
                 & $\Gamma$=210~MeV          & 7$\pm$15${^\dagger}$  & $-$5$\pm$25${^\dagger}$\\
                 & $\Gamma _{\pi}/\Gamma$=0.09&    1$\pm$15  & $-$29$\pm$61 \\
\colrule
$N(1520)D_{13}$  & $W_{R}$=1515~MeV          & $-$47$\pm$2   & $-$125$\pm$2  \\
                 & $\Gamma$=104~MeV          & $-$67$\pm$4   & $-$112$\pm$3  \\
                 & $\Gamma _{\pi}/\Gamma$=0.63&$-$59$\pm$9   & $-$139$\pm$11 \\
\colrule
$N(1675)D_{15}$  & $W_{R}$=1674~MeV          & $-$42$\pm$2   &  $-$60$\pm$2  \\
                 & $\Gamma$=147~MeV          & $-$50$\pm$4   &  $-$71$\pm$5  \\
                 & $\Gamma _{\pi}/\Gamma$=0.39& $-$43$\pm$12 &  $-$58$\pm$13 \\
\colrule
$N(1680)F_{15}$  & $W_{R}$=1680~MeV          &50$\pm$4       & $-$47$\pm$2   \\
                 & $\Gamma$=128~MeV          &29$\pm$6       & $-$58$\pm$9   \\
                 & $\Gamma _{\pi}/\Gamma$=0.70&29$\pm$10     & $-$33$\pm$9   \\
\colrule
\end{tabular}
\end{table*}

\begin{thebibliography}{99}
\bibitem{note}
A more model-independent approach would have the comparison of
pole residues. However, a comparison with the PDG and previous
analyses would not be possible.
\end{thebibliography}
\end{document}